\documentclass[aps,prd,twocolumn,showpacs,preprintnumbers,amsmath,amssymb]{revtex4}

\usepackage{graphicx}
\usepackage{dcolumn}
\usepackage{bm}
\usepackage{epstopdf}

\newcounter{ichi}
\newcounter{ni}
\begin{document}

\title{Neutrino Background Flux from Sources of Ultrahigh-Energy Cosmic-Ray Nuclei}

\author{Kohta Murase$^{1,2}$}
\author{John F. Beacom$^{2,3,4}$}
\affiliation{%
$^{1}$Yukawa Institute for Theoretical Physics, Kyoto University,
Kyoto, 606-8502, Japan\\
$^{2}$
CCAPP, The Ohio State University, 
Columbus, OH 43210, USA\\
$^{3}$
Department of Physics, The Ohio State University, 
Columbus, OH 43210, USA\\
$^{4}$
Department of Astronomy, The Ohio State University, 
Columbus, OH 43210, USA
}%

\date{March 25, 2010}
                        
\begin{abstract}
Motivated by Pierre Auger Observatory results favoring a heavy nuclear composition for ultrahigh-energy (UHE) cosmic rays, we investigate implications for the cumulative neutrino background.
The requirement that nuclei not be photodisintegrated constrains their interactions in sources, therefore limiting neutrino production via photomeson interactions.
Assuming a $dN_{\rm CR}/dE_{\rm CR} \propto E_{\rm CR}^{-2}$ injection spectrum and photodisintegration via the giant dipole resonance, the background flux of neutrinos is lower than $E_\nu^2 \Phi_\nu \sim {10}^{-9}~{\rm GeV}~{\rm cm}^{-2}~{\rm s}^{-1}~{\rm sr}^{-1}$ if UHE nuclei ubiquitously survive in their sources.
This is smaller than the analogous Waxman-Bahcall flux for UHE protons by about one order of magnitude, and is below the projected IceCube sensitivity.
If IceCube detects a neutrino background, it could be due to other sources, e.g., hadronuclear interactions of lower-energy cosmic rays; if it does not, this supports our strong restrictions on the properties of sources of UHE nuclei.
\end{abstract}



\maketitle


\section{Introduction}

The much-anticipated era of high-energy neutrino astronomy seems near~\cite{LM00, HH02}. 
The IceCube detector at the South Pole is nearing completion~\cite{Ahr+04}, and the comparable KM3Net detector in the Mediterranean is being planned~\cite{Kat06}. 
These and higher-energy neutrino detectors, e.g., ANITA~\cite{ANITA}, are expected to reveal unseen aspects of the extreme universe.

One of the main goals is to identify the sources of the cosmic rays, a long-standing mystery. 
While cosmic rays below the knee at $\sim 10^{15.5}$ eV are likely produced by Galactic supernovae, those at higher energies have less certain origins.
There is special interest in ultrahigh-energy cosmic rays (UHECRs)~\cite{Der07}, which have energies above the ankle at $\sim {10}^{18.5}$~eV and which are almost certainly from extragalactic sources. 
Plausible accelerators include active galactic nuclei (AGN)~\cite{RB93,NMA95}, gamma-ray bursts (GRBs)~\cite{Wax95, MINN08}, newly born magnetars~\cite{Aro03} and clusters of galaxies~\cite{NMA95,KRJ96}.
Neutrinos and gamma rays will be important diagnostics of UHECRs, either directly, by pointing to nearby sources, or indirectly, by the levels of the cumulative background fluxes from all sources. 

For the usually-assumed possibility that the UHECRs are protons, there is a large literature
on neutrino production through photomeson interactions inside (e.g., for AGN~\cite{RM98,AD01}, GRBs~\cite{RM98, WB97}, newly born magnetars~\cite{MMZ09}, and clusters~\cite{CB98}) or outside (e.g., via the Greisen, Zatsepin, and Kuzmin process~\cite{BZ69}
or a lower-energy variant~\cite{EK10}) sources. 
Due to large model uncertainties, more general arguments are useful. 
The Waxman and Bahcall (WB)~\cite{WB98} and the Mannheim, Protheroe and Rachen (MPR)~\cite{MPR01} upper bounds on the neutrino background follow from an assumption that the UHECR sources are at least semi-transparent to photomeson interactions, i.e., that each accelerated proton loses at most $\sim 1/2$ of its energy via this process before escape; the details are discussed below. 
These fluxes define reasonable landmarks to assess the sensitivity of neutrino telescopes (we use ``landmark" instead of ``bound" to emphasize that this is a nominal scale instead of an observational bound).

Observations of UHECRs have recently been greatly improved by the High-Resolution Fly's Eye (HiRes) and the Pierre Auger Observatory (PAO). 
Both report a spectrum cutoff at $\sim 60 \times 10^{18}$ eV~\cite{HiR08, PAO08, PAO09s}.  For UHE protons, this is consistent with attenuation due to photomeson interactions with the cosmic microwave background~\cite{GZK66}. 
For UHE heavy nuclei such as iron, as in models in Refs.~\cite{PAO09s,All+05,ABG09}, it is consistent with attenuation due to photodisintegration interactions with the cosmic infrared background~\cite{PAO09s,All+05}. 

Surprising new results suggest that UHECRs may be nuclei instead of protons.
The UHECR composition is probed by the average depth of shower maximum, $X_{\rm max}$, and the r.m.s.\ fluctuations around it, $\delta X_{\rm max}$; while both are subject to uncertainties in the
hadronic models, these are much less for $\delta X_{\rm max}$.
HiRes data on $X_{\rm max}$ favor a proton composition~\cite{HiR09}.
However, with the larger PAO data set, and results on both $X_{\rm max}$ and $\delta X_{\rm max}$, a heavier nuclear composition is favored~\cite{PAO09c}.

We derive new results for the neutrino background due to UHECR sources,
taking into account that the PAO results would require that UHE nuclei
survive photodisintegration interactions in their sources. 
Our landmarks for the neutrino background due to UHE nuclei are significantly lower than the analogous WB and MPR landmarks for UHE protons. 
For all these landmarks, neutrinos are produced by photomeson interactions in the sources.
The difference arises because the requirement that nuclei survive photodisintegration
strongly limits the density of these target photons. 
We explore the conditions that set this landmark, as well as the caveats that apply to its use.


\section{Review of Neutrino Landmarks for the UHE Proton Case}

UHE protons may have photomeson and Bethe-Heitler pair-production interactions with radiation fields in sources. 
Landmarks for the neutrino and gamma-ray background fluxes can be obtained in relation to the cosmic-ray flux.  Key inputs are the typical number of interactions for escaping cosmic rays and the normalization of their injection rate, which depends on their spectrum. 
Photomeson interactions near threshold produce single pions, and charged and neutral pion decays produce neutrinos and gamma rays, respectively. 

The WB landmark for the neutrino background from UHE proton sources~\cite{WB98} is based on three assumptions:
{\bf (a)} the effective optical depth for photomeson interactions is taken to the formal limit of unity for semi-transparent sources, {\bf (b)} the injected cosmic-ray spectrum is $dN_p/dE_p \propto E_p^{-2}$, and {\bf (c)} magnetic fields in the Universe do not affect the observed flux of extragalactic cosmic rays, i.e., magnetic confinement does not change the observed cosmic-ray spectrum.
Assumption (c) becomes most relevant when assumption (b) is altered or not used, as for the more general but higher MPR landmark, which was constructed with constraints on the observed cosmic-ray flux, and allows other classes of cosmic-ray sources below and above ${10}^{19}$~eV; in this case, assumption (a) basically corresponds to the condition that neutrons freely escape from sources~\cite{MPR01}.

For these landmarks, sources are assumed to be semi-transparent for all loss processes, and $f_{p \gamma} \approx t_{\rm int}/t_{p \gamma}$ is the effective optical depth for photomeson interactions, where $t_{p \gamma}$ is the photomeson energy loss time and $t_{\rm int}$ is the interaction duration for cosmic rays.
(For UHECR sources such as GRBs, $t_{\rm int} \approx t_{\rm dyn}$ is used since $t_{\rm dyn} \lesssim t_{\rm  esc}$ is typically assumed; when acceleration is escape-limited or particles leaving the acceleration region propagate in a persistent field, as in clusters, one should use $t_{\rm int} \approx t_{\rm esc}$.)
The WB landmark takes the formal limit $f_{p \gamma} \rightarrow 1$ below.
Strictly speaking, this does not mean $f_{p \gamma} \rightarrow1$, but rather $f_{p \gamma}/(1 - f_p) \rightarrow 1$, where $f_p <1$ is the effective optical depth for all loss processes, since only a fraction $(1 - f_p)$ of produced nucleons can leave the source to contribute to the observed cosmic-ray flux.

For comparison to our results, we reproduce the WB landmark by using assumptions (a) and (b).
We take the energy injection rate of UHE protons to be $E_{p}^2 d \dot{N}_p / d E_p \approx 0.6 \times {10}^{44}~{\rm erg}~{\rm Mpc}^{-3}~{\rm yr}^{-1}$ at ${10}^{19}$~eV for a $dN_p/dE_p \propto E_p^{-2}$ spectrum, consistent with recent PAO results~\cite{KBW09}, as well as with the results from earlier experiments used in Refs.~\cite{WB98, BW03}. 
Though the injection spectrum shape may be different, as discussed in Ref.~\cite{MPR01}, this choice is reasonable for demonstration purposes.
The WB landmark is obtained from
\begin{equation}
E_{\nu}^2 \Phi_{\nu} \approx
\frac{1}{4}  f_{p \gamma}\frac{c t_H}{4 \pi} E_{p}^2 \frac{d \dot{N}_{p}}{d E_{p}}.
\label{eq:neuA}
\end{equation}
Roughly speaking, about half of produced pions are charged, and muon neutrinos ($\nu_{\mu}+\bar{\nu}_{\mu}$) carry about half the pion energy. 
For cosmological parameters $\Omega_m = 0.3$, $\Omega_{\Lambda} = 0.7$, and $H_0 = 70~{\rm km}~{\rm s}^{-1}~{\rm Mpc}^{-1}$, then $t_H \approx 13.5$~Gyr and the formal limit of $f_{p \gamma} \rightarrow 1$ in Eq.~(\ref{eq:neuA}) gives 
\begin{equation}
E_{\nu}^2 \Phi_{\nu} \simeq
1.0 \times {10}^{-8} f_z~{\rm GeV}~{\rm cm}^{-2}~{\rm s}^{-1}~{\rm sr}^{-1}, 
\end{equation}
where $f_z$ is the redshift evolution factor defined in Eq.~(5) of Ref.~\cite{WB98}, and no-evolution and fast-evolution cases correspond to $f_z \approx 0.6$ and $f_z \approx 3$, respectively. 
(Here, as in Refs.~\cite{MINN08,MMZ09}, we use $f_z$, although $\xi_z$ is used in Ref.~\cite{WB98}.)
As long as the relevant assumptions hold for all sources, the WB and MPR fluxes give reasonably optimistic landmarks for the neutrino background, useful for comparing to the sensitivity of neutrino telescopes.
Here and below, neutrino mixing is neglected, to allow direct comparison to previous work, and because the detectors are not sensitive only to muon neutrinos.


\section{Results on Neutrino Landmarks for the UHE Nuclei Case}

If the UHECRs are nuclei instead of protons, the above landmark fluxes can be applicable (for the same requirement on optical depth for photomeson interactions, the UHECR energy range probed is $\sim A$ times higher).

However, for nuclei in radiation fields, the photodisintegration process is even more important than the photomeson process~\cite{Ste69, PSB76}.
We derive a new landmark for the neutrino background by using a new assumption {\bf (a')} that UHECR nuclei survive photodisintegration in their sources and escape without losing their energy; this is more stringent than assumption (a).
As long as the conditions and appropriateness of the underlying assumptions are common in the sources, our nucleus-survival landmarks will work as indicative upper bounds, much like the WB landmark.

For isotropic target photon fields, the photodisintegration interaction time $t_{A \gamma}$ is given by~\cite{Ste69}
\begin{equation}
t^{-1}_{A \gamma}(\varepsilon _{A}) =
\frac{c}{2{\gamma}^{2}_{A}} 
\int_{\bar{\varepsilon}_{\rm th}}^{\infty} \! \! \! 
d\bar{\varepsilon} \, {\sigma}_{A \gamma}(\bar{\varepsilon}) 
\bar{\varepsilon} \int_{\bar{\varepsilon}/2{\gamma}_{A}}^{\infty} 
\! \! \! \! \! \! \! \! \! d \varepsilon \, {\varepsilon}^{-2} 
\frac{dn}{d\varepsilon},
\end{equation}
where $\sigma_{A \gamma}$ is the photodisintegration total cross section, $\bar{\varepsilon}$ is the photon energy in the nucleus rest frame, $\bar{\varepsilon}_{\rm th}$ is the threshold energy, and $\varepsilon_A = \gamma_A m_A c^2$ is the nucleus energy in the source frame (or the comoving frame if the source is moving).
Near threshold, photodisintegration occurs via the giant dipole resonance (GDR), a collective vibration of nucleons. 
At increasingly higher energies, quasi-deuteron emission, baryon resonances, and fragmentation become more relevant.
In many astrophysical situations, the GDR mode is dominant, due to the falling spectra of cosmic rays and target photons, and we focus on such cases. 
In addition to assumptions (a') and (b), this is an assumption required to obtain our landmark fluxes.  
Possible non-GDR effects are discussed later. 


\subsection{Condition on Photodisintegration Optical Depth}

To calculate the photodisintegration rate, we assume that the target photon spectrum is a power-law, $dn/d\varepsilon = n_0 {(\varepsilon/\varepsilon_0)}^{-\alpha}$, as typically expected for sources such as GRBs and AGN. 
If the photon spectrum is sufficiently soft ($\alpha \gtrsim 1$), the photodisintegration cross section can be approximated by the GDR cross section as $\sigma_{A \gamma} \sim \sigma_{\rm  GDR}
\delta(\bar{\varepsilon} - \bar{\varepsilon}_{\rm GDR}) \Delta \bar{\varepsilon}_{\rm GDR}$~\cite{MINN08, WRM08, PMM09}.
Then we have~\cite{MINN08,PMM09}
\begin{eqnarray}
t_{A \gamma}^{-1} \approx \frac{2 \varepsilon_0 n_0}{1+\alpha}
\!\!\! &c& \!\!\! \sigma _{\rm GDR} 
\frac{\Delta \bar{\varepsilon}_{\rm GDR}}{\bar{\varepsilon} _{\rm GDR}}
{\left( \frac{E_{A}}{E_{A 0}} \right)}^{\alpha-1}, 
\label{eq:Ngrate}
\end{eqnarray}
where $\sigma _{\rm{GDR}} \approx 1.45 \times {10}^{-27} A~{\rm cm}^2$ is the GDR cross section,
$\bar{\varepsilon}_{\rm{GDR}} \approx 42.65 A^{-0.21}$~MeV ($0.925 A^{2.433}$~MeV) for $A>4$ $(A \leq 4)$, and $\Delta \bar{\varepsilon}_{\rm GDR} \sim 8$~MeV~\cite{KT93}.
Here, $E_{A 0} = \varepsilon_{A0} \delta \simeq 0.5 m_A c^2 \bar{\varepsilon}_{\rm GDR} \delta/\varepsilon_0$ is the energy of a nucleus interacting with a photon with $\varepsilon_0$, where $\delta$ is the Doppler factor that should be taken into account if the source is moving.
The optical depth for photodisintegration is given by $\tau_{A \gamma} \approx t_{\rm int}/t_{A \gamma}$. 

As noted above, photomeson production occurs on nuclei as well as on nucleons, leading to the production of neutrinos~\cite{Rac96}.
This happens when the energy of a target photon exceeds the pion production threshold in the rest frame of a nucleus. 
The cross section is $\sigma_{\rm mes} \sim \sigma_{p\gamma} A$ (neglecting shadowing) and the energy fraction carried by pions is $\kappa_{\rm mes} \sim \kappa_{p  \gamma}/A$ (treating the other nucleons as spectators).
We avoid a more detailed treatment of photomeson production in nuclei, for which the difficulties may be unnecessary for many astrophysical applications.   
Using the $\Delta$-resonance approximation for the same target photon field as for photodisintegration, the photomeson energy loss rate is~\cite{MINN08,WB97}
\begin{eqnarray}
t_{\rm mes}^{-1} \approx \frac{2 \varepsilon_0 n_0}{1+\alpha}
\!\!\! &c& \!\!\! \sigma _{\Delta} 
\kappa_{\Delta }
\frac{\Delta \bar{\varepsilon}_{\Delta}}{\bar{\varepsilon} _{\rm \Delta}}
{\left( \frac{E_{A}}{E_{A 0}^{(\rm mes)}} \right)}^{\alpha-1}, 
\label{eq:pgrate}
\end{eqnarray}
where $\sigma _{\Delta} \approx 4.4 \times {10}^{-28}~{\rm cm}^2$, $\bar{\varepsilon}_{\Delta} \approx 0.34$~GeV,  $\Delta \bar{\varepsilon}_{\Delta} \sim 0.2$ GeV, and ${\kappa}_{\Delta} \sim 0.2$, and these values are taken from the photomeson production process~\cite{WB97}.
Here, $E_{A 0}^{(\rm mes)} = \varepsilon_{A0}^{(\rm mes)} \delta \simeq 0.5 m_A c^2 \bar{\varepsilon}_{\Delta} \delta/\varepsilon_0 \simeq 19 {(A/56)}^{0.21} E_{A 0}$. 
From Eq.~(\ref{eq:pgrate}), we expect $t_{\rm mes} (E_{A 0}^{(\rm mes)}) \sim t_{p \gamma}  (E_{p 0})$, where $E_{p 0} \simeq 0.5 m_p c^2 \bar{\varepsilon}_{\Delta} \delta/\varepsilon_0 \simeq 0.33  {(A/56)}^{-0.79} E_{A 0}$. 
Hence we obtain $f_{\rm mes} (E_{A 0}^{(\rm mes)}) \sim f_{p \gamma} (E_{p 0})$, as long as $t_{\rm int}$ is the same for nuclei and protons.
This means that the requirement of nucleus-survival limits the flux of photomeson neutrinos from protons as well as that from nuclei.
Then, from Eqs.~(\ref{eq:Ngrate}) and (\ref{eq:pgrate}), we have 
\begin{equation}
f_{\rm mes} (E_{A 0}^{(\rm mes)}) 
\approx \frac{\sigma_{\Delta} \kappa_{\Delta}}{\sigma_{\rm GDR}} \frac{\Delta \bar{\varepsilon}_{\Delta}}{\Delta \bar{\varepsilon}_{\rm GDR}}  \frac{\bar{\varepsilon}_{\rm GDR}}{\bar{\varepsilon}_{\Delta}}  \tau_{A \gamma} (E_{A 0}).
\end{equation}
This is essentially the same as Eq.~(16) in Ref.~\cite{MINN08} (but note that $\tau_{A \gamma}$ in this work was defined as $f_{N \gamma}$ there). 
It is the relation between photodisintegration and photomeson production rates, which depends only on fundamental quantities such as the cross section, though it is justified only when the resonance approximations are valid. 
The requirement of ``complete" nucleus-survival at arbitrary energies (up to the maximum energy), i.e., $\tau_{A \gamma} < 1$, then leads to~\cite{MINN08}
\begin{equation}
f_{\rm mes} \sim f_{p \gamma} \lesssim 1.5 \times{10}^{-3} {(A/56)}^{-1.21},
\label{eq:limitA}
\end{equation}
where cases with $A>4$ are considered. (Hereafter we show cases with $A > 4$ only, since it is straightforward to derive expressions for $A \le 4$.) 
For comparison, recall that the WB flux is formally set by $f_{p \gamma} < 1$ in Eq.~({\ref{eq:neuA}}).

\begin{figure}[t]
\includegraphics[width=0.95\linewidth]{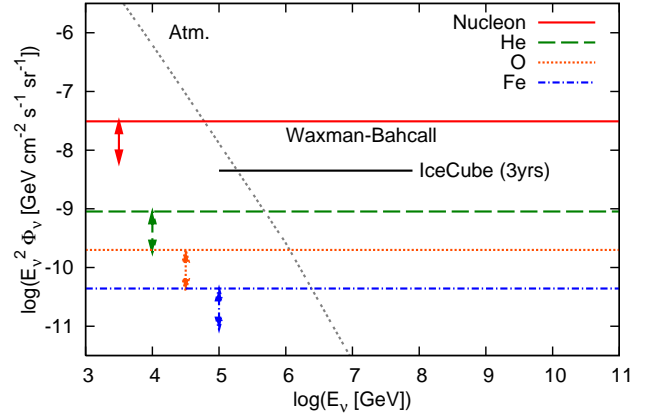}
\caption{\small{\label{MB10bFig1}
Landmarks for the neutrino background from UHECR (injected as $dN_{\rm CR}/dE_{\rm CR} \propto E_{\rm CR}^{-2}$) photomeson interactions in sources, compared to the projected IceCube sensitivity and the atmospheric neutrino background.
The Waxman-Bahcall line for protons is formally set by the upper bound on photomeson interactions in sources ($f_{p \gamma} < 1$).
Our lines for nuclei are set by the upper bounds on photodisintegration interactions in sources ($\tau_{A \gamma} < 1$).
Fast redshift evolution is used, and arrows indicate the change for
no-evolution.
}}
\end{figure}

Similarly to Eq.~(\ref{eq:neuA}), the background flux is written as 
\begin{equation}
E_{\nu}^2 \Phi_{\nu} \approx
\frac{1}{4} f_{\rm mes} \frac{c t_H}{4 \pi} E_{A}^2 \frac{d \dot{N}_{A}}{d E_{A}}.
\label{eq:neuB}
\end{equation}
As a result, for a $E_{\rm CR}^{-2}$ spectrum, the neutrino $(\nu_\mu + \bar{\nu}_\mu)$ background flux is
\begin{equation}
E_{\nu}^2 \Phi_{\nu} \lesssim 1.5 \times {10}^{-11} f_z {(A/56)}^{-1.21}~{\rm GeV}~{\rm cm}^{-2}~{\rm s}^{-1}~{\rm sr}^{-1}. 
\label{eq:boundA}
\end{equation} 
The resulting landmarks are shown in Fig.~\ref{MB10bFig1}, where they are compared to the projected IceCube three-year sensitivity for a $E_{\nu}^{-2}$ spectrum~\cite{Ahr+04}, as well as the estimated atmospheric neutrino background (taken from Ref.~\cite{LM00}, with the spectrum assumed to be $\propto E_{\nu}^{-3}$ above 1 PeV).
The small background fluxes, which will be hard for IceCube and KM3Net to detect, follow from the strong upper limit on interactions with the radiation field required so that all UHECR sources satisfy $\tau_{A \gamma} < 1$.
It is particularly strong when an iron-like composition is assumed as an explanation of the PAO data, as in Refs.~\cite{PAO09s,ABG09}.

Although we have discussed only neutrinos from pion decay, they are also produced by neutron decay following photodisintegration.
However, these neutrinos give lower background fluxes.
The typical neutrino energy in the neutron rest frame is $\sim 0.48$~MeV, and $\tau_{A \gamma} <1$ gives $E_\nu^2 \Phi_{\nu} \lesssim 1.9 \times {10}^{-13} f_z  {(A/56)}^{-1.21}~{\rm GeV}~{\rm cm}^{-2}~{\rm s}^{-1}~{\rm sr}^{-1}$ for electron antineutrinos. 


\subsection{Condition on Photodisintegration Effective Optical Depth}

The PAO composition results are still uncertain, and it is possible that the composition is mixed rather than iron-like.
Also, perhaps a moderate fraction of nuclei undergo photodisintegration interactions in their sources, such that the requirement $\tau_{A \gamma} < 1$ might be too strong.
Instead of this, it would be more conservative to define a condition on the photodisintegration energy loss time $t_{\rm dis}$ for nuclei of initial mass $A$. 

After a heavy nucleus with $A$ (e.g., iron) experiences one photodisintegration interaction via the GDR, the atomic number is $A-1$, which is still heavy.
For the first interaction, the fractional nuclear energy loss, i.e., the inelasticity, is roughly $\kappa_{\rm GDR} \sim 1/A$ around the GDR resonance (since $\gamma_A$ is conserved before and after single-nucleon emission by the GDR)~\cite{PSB76}.
The photodisintegration energy loss time is roughly estimated by multiplying Eq.~(\ref{eq:Ngrate}) by $\kappa_{\rm GDR}$ (or one can evaluate it numerically in a somewhat different
manner~\cite{All+05}).
Then, the more conservative requirement of nucleus-survival is that the effective (energy-loss) photodisintegration optical depth is smaller than unity, i.e., $f_{A \gamma} \approx t_{\rm int}/t_{\rm dis} \sim t_{\rm int} \kappa_{\rm GDR} /t_{\rm A \gamma} <1$.
Then, instead of Eq.~(\ref{eq:limitA}), we have
\begin{equation}
f_{\rm mes} \sim f_{p \gamma} \lesssim 8.2 \times{10}^{-2} {(A/56)}^{-0.21}.
\label{eq:limitB}
\end{equation}
This is larger than that in the previous subsection since some photodisintegration is now allowed.

\begin{figure}[t]
\includegraphics[width=0.95\linewidth]{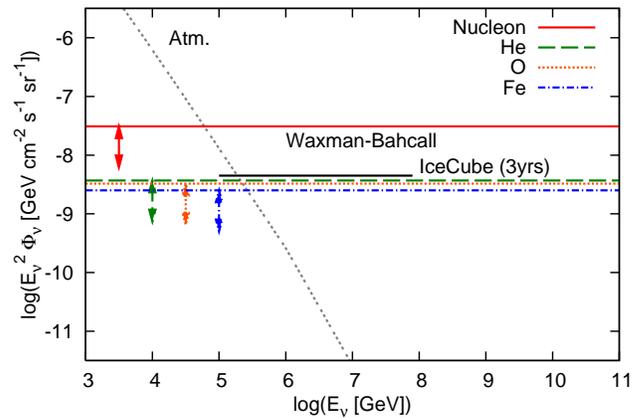}
\caption{\small{\label{MB10bFig2}
The same as Fig.~1, but the photodistintegration bound is defined
instead by $f_{A \gamma} < 1$.
}}
\end{figure}

The corresponding nucleus-survival landmark for the neutrino background is analogous
to Eq.~(\ref{eq:neuB}). 
However, when nucleons are ejected from nuclei via the GDR, both the nuclei themselves and the ejected nucleons produce neutrinos via photomeson interactions.
Instead of Eq.~(\ref{eq:neuB}), in more generality, we have
\begin{eqnarray}
E_{\nu}^2 \Phi_{\nu} &\approx& \frac{1}{4}  \frac{c t_H}{4 \pi} \left[ f_{p \gamma}(E_{A}/A) f_{A \gamma} (E_A)
 \right. 
\nonumber \\ 
&+& \left. 
f_{\rm mes}(E_A) (1-f_{A \gamma}(E_A)) 
\right] 
E_{A}^2 \frac{d \dot{N}_{A}}{d E_{A}},
\label{eq:neuC}
\end{eqnarray}
where we have still assumed $f_{A \gamma}<1$. 
However, because $f_{p \gamma}(E_A/A) \sim f_{\rm mes}(E_A)$, this becomes the same as Eq.~(\ref{eq:neuB}). 
Hence, similarly to Eq.~(\ref{eq:boundA}), the neutrino $(\nu_\mu + \bar{\nu}_\mu)$ background flux is 
\begin{equation}
E_{\nu}^2 \Phi_{\nu} \lesssim 8.4 \times {10}^{-10} f_z {(A/56)}^{-0.21}~{\rm GeV}~{\rm cm}^{-2}~{\rm s}^{-1}~{\rm sr}^{-1}, 
\label{eq:boundB}
\end{equation}
which is still lower than the WB landmark by one order of magnitude.
The near-$A$-independence of this result is a consequence of the fact that $\sigma_{\rm GDR} \, \kappa_{\rm GDR} \sim A (1/A) \sim 1$; in the previous subsection, the term $\kappa_{\rm GDR}$ was not included.
The results are shown in Fig.~\ref{MB10bFig2}.

The neutrino background from nuclei accelerators was briefly considered in Ref.~\cite{Anc+08}, where it was argued that this flux is much smaller than the WB flux.
Our work is different, since we quantitatively take into account the nucleus-survival condition, showing that it is crucial to constrain properties of the sources, and that it leads to a small but appreciable
neutrino flux.

Similarly to Eq.~(\ref{eq:boundB}), the landmark for neutrinos from neutron decay following photodisintegration can be obtained; the condition $f_{A \gamma}<1$ leads to $E_\nu^2 \Phi_{\nu} \lesssim {10}^{-11} f_z  {(A/56)}^{-0.21}~{\rm GeV}~{\rm cm}^{-2}~{\rm s}^{-1}~{\rm sr}^{-1}$ for electron antineutrinos. 

\subsection{Dependence on Spectral Index}

The nucleus-survival landmarks expressed in Eqs.~(\ref{eq:boundA}) and (\ref{eq:boundB}) were
derived for a $E_{\rm CR}^{-2}$ spectrum. 
Different indices are allowed from UHECR observations, depending on source evolution models.
Here, modifying assumption (b), we consider the case where
$\frac{dN_{\rm CR}}{dE_{\rm CR}} \equiv \Sigma_{A \geq 1} \left( \frac{dN_{A}}{dE_{A}} \right) = \Sigma_{A \geq 1} \left( y_A \frac{dN_{\rm CR}}{dE_{\rm CR}} \right)$ with $ \frac{dN_{A}}{dE_{A}} \propto E_{A}^{-s}$.
Here, $y_A$ is the fraction of nuclei with mass $A$.
As an example, assuming a two-component case,
Eq.~(\ref{eq:neuA}) is replaced by
\begin{eqnarray}
E_{\nu}^2 \Phi_{\nu} \sim \frac{1}{4}  \frac{c t_H}{4 \pi} \left[ f_{p \gamma} E_{p}^2 \frac{d \dot{N}_{p}}{d E_{p}}  \right. 
+ \left. 
f_{\rm mes} E_{A}^2 \frac{d \dot{N}_{A}}{d E_{A}}  
\right],
\label{eq:neuD}
\end{eqnarray}
where we have used $f_{p \gamma} (E_A/A) \sim f_{\rm mes}(E_A)$.
For the UHECR energy injection rate at ${10}^{19}$~eV, we use $E_{\rm CR}^2 d N_{\rm CR} / d E_{\rm CR} = 0.6 \times (s-1){10}^{44}~{\rm erg}~{\rm Mpc}^{-3}~{\rm yr}^{-1}$~\cite{KBW09}.

\begin{figure}[t]
\includegraphics[width=0.95\linewidth]{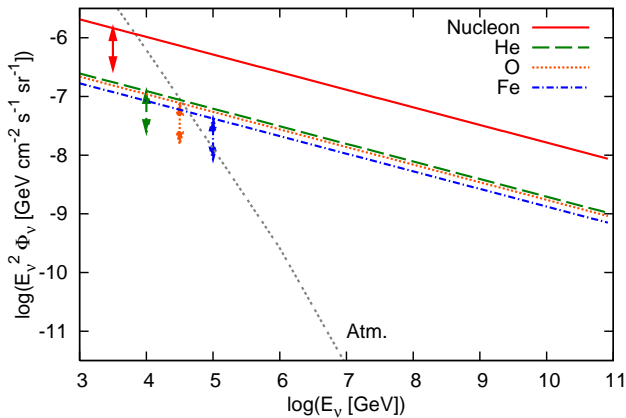}
\caption{\small{\label{MB10bFig3}
The same as Fig.~2, but the UHECR are
injected as $dN_{\rm CR}/dE_{\rm CR} \propto E_{\rm CR}^{-2.3}$.
}}
\end{figure}

To set landmarks, we take only the larger of the two terms above (one for protons, one for nuclei). 
A neutrino of energy $E_\nu$ can be produced by a proton of energy $E_p \approx 20 E_{\nu}$ or a nucleus of energy $E_{A} \sim 20 A E_{\nu}$; the cosmic-ray energy flux is smaller (larger) at the higher energy for $s>2$ $(s<2)$.  For $s >2$, Eq.~(\ref{eq:boundB}) is replaced by 
\begin{eqnarray}
E_{\nu}^2 \Phi_{\nu} \lesssim 8.4 &\times& {10}^{-10}~{\rm GeV}~{\rm cm}^{-2}~{\rm s}^{-1}~{\rm sr}^{-1} \nonumber \\
&\times& f_z {(A/56)}^{-0.21} (s-1) E_{\nu,17.7}^{2-s}.
\label{eq:boundC}
\end{eqnarray}
For $s<2$, Eq.~(\ref{eq:boundB}) is replaced by
\begin{eqnarray}
E_{\nu}^2 \Phi_{\nu} \lesssim 8.4 &\times& {10}^{-10}~{\rm GeV}~{\rm cm}^{-2}~{\rm s}^{-1}~{\rm sr}^{-1} \nonumber \\
&\times& f_z {(A/56)}^{1.79-s} (s-1) E_{\nu,15.95}^{2-s}.
\label{eq:boundD}
\end{eqnarray}
Results for an example with $s>2$ are shown in Fig.~\ref{MB10bFig3}.

\subsection{Dependence on Composition}

Our results in Eqs.~(\ref{eq:boundA}), (\ref{eq:boundB}), (\ref{eq:boundC}) and (\ref{eq:boundD}) do not depend on the composition itself. 
They do have $A$-dependence, which comes from which nucleus is adopted for the nucleus-survival condition. 
Generally speaking, the requirement $f_{A \gamma} <1$ (or $\tau_{A\gamma}<1$) leads to composition-dependent landmarks, since Eq.~(\ref{eq:neuD}) implicitly includes $y_A$. 
However, we can define our landmark by just the largest of the composition-dependent terms, as done in deriving Eqs.~(\ref{eq:boundC}) and (\ref{eq:boundD}). 
The energy of a nucleus producing a neutrino with energy $E_{\nu}$ is higher than that of a proton producing a neutrino with energy $E_\nu$.
For $s>2$, the first term in Eq.~(\ref{eq:neuD}) is more important.
Thus, for the same $y_A$, more neutrinos with $E_{\nu}$ come from protons than from nuclei.
For $s<2$, the second term is more important, and the situation is reversed. 
For $s=2$, because $f_{p \gamma}(E_A/A) \sim f_{\rm mes}(E_A)$, the landmark neutrino flux is already independent of the composition.


\subsection{Discussion of Applicability}

While our results are general, they must be accompanied by some caveats.

First, our arguments are valid only when the photodisintegration and photomeson interactions are both governed by resonances (the ratio of our nucleus-survival landmark and the WB landmark essentially follows from the relative properties of the GDR resonance and the $\Delta$ resonance). 
In principle, non-resonance effects could be important over a broad energy range. 
For example, in the high-energy limit, photodisintegration is governed by fragmentation, where many pions and nucleons are produced, and then one would not expect a significant difference between $f_{\rm mes}$ (or $f_{p \gamma}$) and $f_{A \gamma}$. 
However, in many astrophysical situations, both the photodisintegration and photomeson processes are well-described by resonance approximations as long as the target photon spectrum is soft enough~\cite{MINN08,MPR01,PMM09,Rac96}. 
For a power-law photon spectrum with $\alpha \sim 2$, the resonance approximations are good~\cite{MINN08,MPR01,PMM09}, and our landmarks are valid; if  $\alpha \sim 1$, non-resonance effects are moderately important \cite{MINN08,MPR01,Rac96,MN06,Muc+99}. 
For a black-body photon spectrum with temperature $T$, the energy loss rate is maximal around $E_{A 0} \simeq 0.5 m_A c^2 \bar{\varepsilon}_{\rm GDR} \delta/kT$, so that one only has to consider the nucleus-survival condition at this energy as long as the injection energy of nuclei is lower than $E_{A 0}$.
Thus, in practice, our results would be valid when the targets are radiation fields with sufficiently soft spectra, as considered here.

On the other hand, for hadronuclear processes, including the $pp$ interaction, where the non-resonant region is crucial except near the pion-production threshold, our results would not be valid.
At high energies, this cross section is $\sigma_{Ap} \sim 5 \times {10}^{-26} A^{2/3}~{\rm cm}^2$ (in the shadowing limit), where spallation, fragmentation and meson production occur.  Detailed studies would be required for nucleus-survival landmarks in such cases, and are beyond the scope of this work.  

Second, in deriving Eqs.~(\ref{eq:limitA}) and (\ref{eq:limitB}), we implicitly assumed that $t_{\rm int}$ is the same for nucleons and nuclei.  For transients such as GRBs, $t_{\rm int} \approx t_{\rm dyn}$ is expected.  But for persistent sources, one may expect $t_{\rm int} \approx t_{\rm esc}$, and the escape time may be different between the two. 
In addition, particle escape could be related to another problem. 
Since the landmarks are normalized by the UHECRs, assumption (a') implies that all the UHECRs accelerated in the sources contribute to the observed UHECR flux (i.e., the effective optical depth for all loss processes $f_{A} \ll 1$). 
This might be true, especially at the highest energies; for example, for GRBs we may expect $t_{\rm esc} \sim t_{\rm dyn} \sim t_{\rm acc} \lesssim t_{\rm cool}$~\cite{MINN08}. 
But, if the escape time is too long, cosmic rays might lose their energies before escape, via adiabatic or radiative cooling, and so on (for UHE nuclei sources, we cannot use the neutron escape mechanism, which may work in UHE proton sources).
This potentially allows hidden accelerators, where a significant fraction of accelerated cosmic rays would not contribute to the observed cosmic-ray flux; this would then produce more neutrinos than expected.
Particle escape is one of the open problems in particle acceleration theories and we avoid further considerations for simplicity. 

Third, assumption (a') might be too strong in the sense that the nucleus-survival condition might not be satisfied ubiquitously.
This condition is sensitive to UHECR source properties,
which are uncertain, and which may have a large diversity.
As an example, suppose that
the nucleus-survival condition is satisfied for $\sim 9/10$ of sources, but not for $\sim 1/10$ of sources,
and that the latter have high target photon densities, as for the WB flux.
As a result, the neutrino background flux could exceed Eqs.~(\ref{eq:boundB}) or (\ref{eq:boundC}) or (\ref{eq:boundD}).  For this example, taking $s=2$, the flux would be roughly doubled compared to the landmark flux expressed in Eq.~(\ref{eq:boundB}).   

Fourth, landmark fluxes are derived for a specific assumed cosmic-ray injection spectrum.  In more generality, as in Ref.~\cite{MPR01}, ``bounds" could be obtained by comparing with all cosmic-ray data. 
In the case of nuclei, one would need to compare to both spectrum and composition data,
which would require calculating cascades for nuclei inside and outside the sources.  Such detailed calculations are deferred here, because Eq.~(\ref{eq:boundB}) for $s=2$ and Eq.~(\ref{eq:boundC}) for $s > 2$ are enough for demonstration purposes, and the composition data are not yet adequate.
In addition, when assumption (b) is not adopted, we would have to take more care with assumption (c) (see, e.g., Refs.~\cite{Lem05} for effects of cosmic magnetic fields). 
Nuclei are more easily deflected by magnetic fields, so then the landmarks might not be so stringent.


\section{Conclusions}

If the PAO data correctly indicate that UHECRs have a heavy composition, then a significant fraction of nuclei must survive photodisintegration interactions in their sources (further,
some sources must be nearby, since nuclei from distant sources cannot avoid photodisintegration en route).
For probing the density of target photons in sources, nuclei are special compared to protons, since interactions lead to changes in composition and energy, and not just a change in energy.
For UHE protons, the effective requirement $f_{p \gamma}<1$ is theoretical, not observational,
and $f_{p \gamma}>1$ is allowed in principle~\cite{MPR01}.
For UHE nuclei, the requirement $f_{A \gamma} < 1$ in typical sources seems to be observationally required by the PAO results, though see the caveats above.
More detailed discussions must wait for more precise composition results.

We present new, theoretically-indicated landmarks for the cumulative neutrino background, following from the condition of nucleus-survival in UHECR sources. 
If this is satisfied in \textit{all} UHECR sources, then the resulting landmark neutrino flux is at least one order of magnitude smaller than the WB flux for UHE protons.
Detection in IceCube or KM3Net of the neutrino background produced by the photomeson interactions of UHE nuclei in their sources could be challenging.
While not equivalent to experimental bounds, our landmarks are less model-dependent than predictions for specific sources, and will be useful as general probes of UHECR sources and for assessing the sensitivity of neutrino telescopes.
For some specific models in which UHE nuclei survive in their sources, e.g., the various GRB models of Ref.~\cite{MINN08}, the neutrino fluxes are below our landmarks, as expected.
Thus our results indicate reasonably optimistic neutrino background fluxes.

Although our arguments can be applied to the neutrino background produced inside sources of UHE nuclei, they cannot be applied to the cosmogenic neutrino fluxes produced outside these sources.
This is because assumption (a') will not hold; for example, the energy attenuation length of iron is $\sim 100$~Mpc at $\sim {10}^{20}$~eV~\cite{All+05,PSB76}, which implies that UHE nuclei
from very distant sources are significantly disintegrated.
As a result, the fluxes of cosmogenic neutrinos from nuclei strongly depend on the spectral index, maximum energy and source evolution~\cite{Ave+05}.  If the maximum energy is very high, those nuclei are completely disintegrated and the emitted nucleons undergo photomeson interactions, similarly to the case where protons are initially injected. 
However, one expects that such cases would conflict with the PAO composition data.
Detailed works suggest that the cosmogenic neutrino flux in the nuclear case would be much smaller than that in the proton case~\cite{Ave+05}.  

As discussed, if a fraction of sources violate assumption (a'), then the neutrino flux can exceed the nucleus-survival landmark.
Future neutrino observations can test this. 
Recent neutrino observations are already almost reaching the WB landmark for ``integral" limits (where power laws are assumed over a few decades)~\cite{Ahr+04}, suggesting that hidden neutrino sources are excluded and that assumption (a) is indeed reasonable. 
Similarly, if assumption (a') is valid, IceCube and KM3Net may not see neutrinos from UHECR sources. 
If IceCube sets a tight limit on the neutrino background, this would suggest that almost all sources do have the low target photon densities required for nucleus-survival. 
If IceCube measures a larger flux, this might still be largely true, though there are other possibilities.
One is that there could be some proton sources or hidden accelerators. 
Another is that neutrinos are also produced via hadronuclear processes, e.g., such as clusters of galaxies~\cite{CB98}.

To reveal the accelerators of UHECRs, detections of gamma rays are also important. 
Especially, signals unique to UHECR accelerators are needed.
For UHE proton sources, there could be UHE pionic gamma rays~\cite{Mur09} and GeV synchrotron gamma rays~\cite{GA05}.
For UHE nuclei sources, there could be TeV-PeV gamma rays from nuclear de-excitation following photodisintegration~\cite{MB10}.

\medskip

{\bf Acknowledgments:}
K.M. thanks K. Ioka for encouraging comments.
K.M. was supported by a Grant-in-Aid from JSPS and by CCAPP. 
J.F.B. is supported by NSF CAREER Grant PHY-0547102. 



\end{document}